\documentstyle[twocolumn,prb,aps,epsf,epsfig]{revtex}
\begin{document}

\title{
Arrays of Josephson junctions in an environment with
vanishing impedance}

\author{M. Aunola, J.J. Toppari and J.P. Pekola}

\address{Dept. of Physics, University of Jyv\"askyl\"a,
P.O. Box 35 (Y5), FIN-40351 Jyv\"askyl\"a, Finland}

\date{\today}
\maketitle

\begin{abstract}
The Hamiltonian operator for an unbiased array of Josephson junctions
with gate voltages is constructed when only Cooper pair tunnelling and
charging effects are taken into account.  The supercurrent through the
system and the pumped current induced by changing the gate voltages
periodically are discussed with an emphasis on the inaccuracies in the
Cooper pair pumping. Renormalisation of the Hamiltonian
operator is used in order to reliably parametrise
the effects due to inhomogeneity in the
array and non-ideal gating sequences. The relatively simple model
yields an explicit, testable prediction based on three
experimentally motivated and determinable parameters.
\end{abstract}

\section{Introduction}

When a potential well propagates  adiabatically along an
electron system that is effectively one-dimensional,
it carries with it additional electron density, and induces
dc electric current through the system. Such a pumping
effect has been studied in small metallic tunnel junctions in the
Coulomb blockade regime.\cite{pot92,kel96}
If the propagation of the potential well is arranged by
phase-shifted gate voltages as in Refs. 1 and 2
and the potential well carries a quantised number $n$
of electrons then the induced current $I$ is related
to the gating frequency $f$ by the fundamental relation $I=-nef$,
where $e=1.602\cdot10^{-19}$ C.
These Coulomb blockade pumps transporting single, normal-state
electrons have reached accuracy suitable for metrological
applications.\cite{kel96,kel99}

Until lately, mainly pumping of single electrons was
studied, but in a recent article\cite{pek99} a quantitative
theory of pumping Cooper pairs in gated one-dimensional
arrays of Josephson junctions was presented. It was
shown that quantum effects render the Cooper pair pump
inaccurate in case of arrays with small junctions thus
explaining the failure to demonstrate accurate pumping in the first
(and so far the only) reported experiment of pumping of Cooper
pairs.\cite{gee91}

In this article, using renormalisation methods,
we generalise the results derived in Ref. 4 for
an unbiased array of Josephson junctions with gate voltages when
only Cooper pair tunnelling and
charging effects are taken into account.
We consider the supercurrent, the higher order corrections for
the pumping inaccuracy in case of homogeneous arrays,
inhomogeneity of the array, and nonideal pumping sequences.
Using three experimentally motivated and determinable parameters
we are able to derive an expression for the pumping inaccuracy
which can be directly compared against experimental results. 

It should be stressed, though,
that the model is simple and neglects many, possibly important features
such as quasiparticle tunnelling, the coupling to the
electromagnetic environment and the dissipative
effects induced by a non-zero bias voltage. On the other hand,
without these simplifications the problem could not be
solved at the moment.
The efficiency of the renormalisation methods in case of Cooper pair
pumps is explained by the considerable
symmetries of the model Hamiltonian.
Renormalisation is widely used in atomic and nuclear
physics,\cite{ell78,hjo98,hjo95,kuo90,lin85} e.g. for
the creation of effective two-body interactions used in
nuclear Shell-Model calculations.

The article is organised as follows. In Secs.~\ref{sec:renorm}
and \ref{sec:cooper} the renormalisation method is explained
and the expressions for the Hamiltonian as well as the charge
transported by pumping are derived, respectively.
In Sec.~\ref{sec:homog} homogeneous arrays are
examined in detail and in Sec.~\ref{sec:inhomog} the
inhomogeneity is introduced. Finally in
in Sec.~\ref{sec:practic} nonideal gating sequences are
considered and the theoretical prediction for the pumping
inaccuracy is explained thus leading to the conclusions in
Sec.~\ref{sec:conclu}.

\section{The renormalisation method}\label{sec:renorm}

\subsection{Few-state dominant systems}

Renormalisation methods  may be applied most effectively
in case of ``few-state dominant'' systems. In such systems
some of basis states are separated from all the others
by an energy $\Delta E$ which is large as compared
to the coupling between any two states in
the system. Thus certain eigenstates
of the full system can be approximated by the
eigenstates of the "few-state dominant" part only.
Renormalisation can be described as a bridge spanning
across the intervening gap.

For a given orthogonal basis $\{\vert l\rangle\}$
the Hamiltonian operator $H$  can be written as
\begin{equation}
H=H_0+V,\ \ H_0\vert l\rangle=\epsilon_l\vert l\rangle,
\ \ l=1,2,\ldots
\end{equation}
where $V$ is the residual interaction defining
the coupling between eigenstates of $H_0$.
The matrix elements of $V$ can often be considerably suppressed by a
proper choice of the basis states.
Explicitly, a part of the Hamiltonian can be called
"$k$-state dominant" if there are $k$ basis states that
satisfy the conditions
\begin{eqnarray}
&\vert \epsilon_l-\epsilon_m\vert\ge \Delta E,&\quad{\rm if}
\ \ l\le k<m,\cr
&\vert V_{nn'}\vert\ll \Delta E,&\quad\ {\rm for\ all}\ n,n'.
\end{eqnarray}
The requirements of ``few-state dominance''
are graphically depicted in Fig.~\ref{fig:first}
showing some of the energy levels for  two  examples of a
$5$-state dominant system. Note that $V_{\rm max}$, the
magnitude of the largest element in $V$, can be much smaller than
the energy spread of the low-lying basis states.

\subsection{The effective interaction}\label{sec:effect}

The
aim of renormalisation is to generate an effective interaction
$\tilde V$ for a small, active space ($P$-space) which yields the same
eigenvalues and eigenstates as the full Hamiltonian operator $H$.
Similarly effective operators can be defined but this
complication can be avoided in case of Cooper pair pumps.
The renormalisation method in the context
of nuclear physics is carefully explained in Refs. 6 and 7.

The effective interaction $\tilde V$ for
a $k$-dimensional $P$-space spanned by
$\{\vert l\rangle\}_{l=1}^k$ can be derived as follows.
One starts from the full Hamiltonian equation
\begin{equation}
H\vert\psi\rangle=E\vert\psi\rangle\label{eq:hami0},
\end{equation}
where $E$ is an eigenvalue,
inserts the expansion $\vert\psi\rangle=
\sum_la_l\vert l\rangle$ in order to obtain the
set of linear equations for the coefficients $\{a_l\}$.
In the first $k$ equations of Hamiltonian the rest of the
equations can be repeatedly applied in the form
\begin{equation}
a_{m(>k)}=\sum_{l=1}^{k}\frac{V_{ml}a_{l}}{E-\epsilon_m}+
\sum_{m'(>k)}\frac{V_{mm'}a_{m'}}{E-\epsilon_m}.\label{renconv}
\end{equation}
The renormalisation eliminates coefficients
$a_{m(>k)}$ and converges certainly if
\begin{equation}
\sum_{m'(>k)}\frac{\vert V_{mm'}\vert}{\vert E-\epsilon_m\vert}<1
\label{rowsum}
\end{equation}
for all states $\vert m\rangle$ with $m>k$.  Inclusion of additional
basis states in the $P$-space improves the convergence by reducing the
number of coefficients that must be eliminated.  Violation of the property
(\ref{rowsum}) does not necessarily imply divergence, but the
convergence should be checked more carefully. 

We define operators $\hat P=
\sum_{l=1}^k\vert l\rangle\langle l\vert$ and $\hat Q=\hat 1-\hat P$
that project onto the $P$-space and the rest of the full space
($Q$-space), respectively. The
renormalised Hamiltonian can then be written as
\begin{equation}
\tilde H=\hat P H_0\hat P+\tilde V.\label{eq:effective}
\end{equation}
where the effective interaction is given by
\begin{equation}
\tilde V=\hat P\left[\sum_{n=0}^{\infty}\left(V\frac{\hat Q}
{E-\hat Q H_0\hat Q}\right)^n\right]V\hat P.
\label{vrenorm}
\end{equation}
When $\tilde V$ converges $E$ is also an eigenvalue of
renormalised Hamiltonian $\tilde H$ which is
manifestly consistent with the full Hamiltonian equation
(\ref{eq:hami0}). But, unless $\tilde V$ can be evaluated
(analytically) up to the infinite order,
the renormalisation should be truncated
at a given order or when preset convergence criteria
are met. This approximate effective interaction
replaces $\tilde V$ in Eq.~(\ref{eq:effective}).

The only  remaining question is how the eigenenergy $E$
should be chosen? The obvious, correct answer is
"Insert $E$, get $E$". The effective Hamiltonian
$\tilde H$ yields, allowing for convergence, $k$ eigenvalues
$\{\tilde E_j\}_{j=1}^k$. By repeatedly inserting
an eigenvalue $\tilde E_j$ into
the definition of the effective interaction $\tilde V$
we obtain the best available approximation for the
eigenvalue $E_{[j]}$ of $H$ as well as the renormalised
eigenstates $\vert\tilde j\rangle$.
A good initial guess is the eigenvalue $E_{j}^{(P)}$
of the pure $P$-space Hamiltonian $\hat P H\hat P$.
We call this choice "individual" because the
renormalisation has to be performed for each eigenstate
in $P$-space separately.

The eigenstates $\{\vert\tilde j\rangle\}$ are not orthogonal in
$P$-space because they correspond to different effective Hamiltonians,
but if they are reexpanded to the full space the orthogonality can be
regained.  However, the requirements for "few-state dominance" ensure
that the important advantage of orthogonality can be used in exchange
for only a small loss in accuracy.  In the "average" choice
renormalisation, valid for "few-state dominant" systems, we use the
average eigenenergy $\bar E=k^{-1}\sum_{j=1}^k\tilde H_{jj}$ in place
of individual eigenvalues. Thus the evaluation the diagonal matrix
elements $\{\tilde V_{jj}\}$ suffices up until the final step of the
iteration when $\tilde H$ has to be diagonalised.

If the self-consistent iteration is not used we obtain
the results directly corresponding to initial guesses.
We refer to these cases as "individual-0" and "average-0"
choices. These choices have been used since they often offer
a more transparent interpretation of the result as well
as much desired analytical results.

Finally we must emphasise that renormalisation and full
diagonalisation in a restricted basis are just two similar although
unidentical approaches to the eigenvalue problem. In renormalisation
the full problem is projected onto a smaller space while
in diagonalisation the problem is truncated by discarding all
basis states outside the restricted basis.

\section{The Cooper pair pump}\label{sec:cooper}

\subsection{General properties}

An array of Josephson junctions with gate voltages,
a Cooper pair pump (CPP), in the Coulomb blockade
regime is an excellent example of a
few-state dominant system. In an earlier paper\cite{pek99} 
the leading component of the inaccuracy for transferred 
charge in a homogeneous Cooper pair pump was derived. 
The higher order corrections to this  leading order result are
relatively insignificant in the immediate pumping regime.
In that article the inaccuracy was evaluated from
the variance of the number of Cooper pairs on an island far
away from the island where most of the charge
transfer occurs.

A crude version of the renormalisation process, amounting
to ``average-0'' choice was also used in Ref. 4 for crosschecking
the results. An analytical  power expansion of the inaccuracy
derived below was known as an "intelligent guess"
but not included in the results.
In addition, an analytical result for the pumped charge
along a circular path for $N=3$ on the gate voltage  plane
was derived by renormalisation.

Figure~\ref{fig:cppump} shows a schematic drawing of
a gated Josephson array of $N$ junctions.
Each  junction has a capacitance $C_k$ and
a Josephson energy $E_{{\rm J},k}$. The phase difference
over the array is $\phi\in[0,2\pi N)$. It may be controlled
by an external bias voltage $V$ over the array according to
the relation
\begin{equation}
\frac{d\phi}{dt}=\frac{(-2e)V}\hbar.
\end{equation}
The oscillation frequency of $\phi$  is approximately
$V$ [$\mu$V\,$]\cdot 0.5$ GHz, but
all calculations in this article are done under an assumption of ideal
zero bias ($V\equiv0$) yielding a constant $\phi$.
Each gate voltage $V_{{\rm g},k}$ binds charge $C_{{\rm g},k}
V_{{\rm g},k}$ on island $k$. In order to (hopefully) transport
exactly one Cooper pair through the array the gate
voltages are operated as depicted in Fig.~\ref{fig:cppump}.
Thus the  cycle consists of $N$ legs and during $k^{\rm th}$
leg we expect one Cooper pair to tunnel through junction $k$.

There are two important energy scales
in a CPP. The first one is the typical Josephson coupling energy
$E_{\rm J}$ related to the Cooper pair tunnelling through the junctions
and the second one is the charging energy $E_{\rm C}$ related
to the charging effects of the small islands between
the junctions. Both are explicitly defined below.
The most important parameter of the model is the
ratio $E_{\rm J}/E_{\rm C}$ which will be denoted by
$\varepsilon_{\rm J}$.

\subsection{The Hamiltonian and supercurrent operators}

When we neglect the quasiparticle tunnelling and other degrees of
freedom the model Hamiltonian is given by
\begin{equation}
H=H_{\rm C}+H_{\rm J}\label{simphami}
\end{equation}
where $H_{\rm C}$ is the charging Hamiltonian and $H_{\rm J}$
describes the Josephson tunnelling of the Cooper pairs.
For an array of $N$ junctions the tunnelling Hamiltonian
has the form
\begin{equation}
H_{\rm J}=-\sum_{k=1}^{N}E_{{\rm J},k}\cos \hat\phi_k
\end{equation}
where $\hat\phi_k$ is the phase difference over the junction $k$,
corresponding to a supercurrent operator
\begin{equation}
I_{{\rm S},k}=\frac{(-2e)E_{{\rm J},k}}\hbar
\sin \hat\phi_k=\frac{-2e}{\hbar}
\frac{\partial H_{\rm J}}{\partial\hat\phi_k}.
\end{equation}
The charging Hamiltonian $H_{\rm C}$ is diagonal in the basis
formed by the charge eigenstates $\vert \vec n\rangle$ where
$\vec n\equiv\{n_1,n_2,\ldots,n_{N-1}\}$ and
$n_i$ is the number of Cooper pairs on each island of the array.
The normalised gate voltages $\vec q\equiv
\{q_1,q_2,\ldots,q_{N-1}\}$ where $q_k=V_{{\rm g},k}C_{{\rm g},k}
/(-2e)$ may be considered as parameters in $H_{\rm C}$. The
diagonal matrix elements are given by the classical
charging energy
\begin{equation}
E_{\vec n}\equiv E_{\rm ch}^{(\vec n,\vec q)}=\sum_{k=1}^N
\frac{Q_k^2}{2C_k}\label{gencharen}
\end{equation}
where $Q_k=(-2e)v_k$ is the charge across the junction $k$.
We define the typical charging energy of the array as
$E_{\rm C}=(2e)^2/2C$ where the "average"
capacitance $C$ is given by
$C=N/\sum_{k=1}^NC_k^{-1}$ so that $C_k=c_kC$.
The condition for ideal biasing yields $\sum_{k=1}^Nv_k/c_k=0$ and
the conservation of charge on each island requires that
\begin{equation}
v_k-v_{k+1}=u_k \label{veeesi}
\end{equation}
where $\vec u=\vec n-\vec q$. The unique solution
satisfying these conditions is given by $v_k=\tilde v_k+y$ where
\begin{equation}
\tilde v_k=\sum_{j=k}^{N-1}u_j-\frac1N\sum_{j=1}^{N-1}ju_j,
\quad y=-\frac1N\sum_{k=1}^N\frac{\tilde v_k}{c_k}.\label{eq:hipsolu}
\end{equation}
By substituting the solution~(\ref{eq:hipsolu}) into
Eq.~(\ref{gencharen}) we find
\begin{equation}
E_{\vec n}=
E_{\rm C}\left[\sum_{k=1}^N\frac{v_k^2}{c_k}
-\frac1N\left(\sum_{k=1}^N\frac{v_k}{c_k}\right)^2\right].
\label{generchar0}
\end{equation}
where $v_k=\tilde v_k+\bar y$ for arbitrary $\bar y$ since the
expression (\ref{generchar0}) is invariant under the transformation
$\{v_k\}\rightarrow \{v_k+\bar y\}$.\cite{end:biasing} This symmetry
can be effectively applied when renormalised matrix elements are
evaluated.

The optimal basis for calculations is $\{\vert\vec n, \phi\rangle\}$,
the basis of charge eigenstates augmented by the total phase
difference over the array $\phi=\sum_{k=1}^{N}\phi_k$ which is
is periodic over $2N\pi$.
The completeness of the basis was shown in case of normal-state electron
systems by Ingold and Nazarov\cite{ing92} who also give the canonical
transformation between variables describing $N$ separate junctions and
variables describing $N-1$ islands and the array as a whole. 
The conjugate phases on islands $\{\theta_k\}$ as well as the average
number of tunnelled Cooper pairs  ${\cal N}=(1/N)\sum_{k=1}^Nm_k$
($m_k$ naturally corresponds to junction $k$) are completely
undefined for the chosen basis states.

From now on we consider only the case when the phase difference
over the array $\phi$ is kept fixed by ideal biasing and
since $\phi$ is a constant of motion for the 
model Hamiltonian (\ref{simphami}) we can explicitly write
\begin{equation}
H=H_{\rm C}(\vec q)-\sum_{\vec n,k=1}^{N}\frac{E_{{\rm J},k}}2(\vert
\vec n+\vec\delta_k\rangle\langle \vec n\vert e^{i\phi/N}+\,{\rm H.c.}\,).
\label{coophami}
\end{equation}
Here the tunnelling vector $\vec\delta_k$ describes the change of
$\vec n$ due to tunnelling of one Cooper pair through the $k^{\rm th}$
junction. The non-zero components of $\vec \delta_k$ are (if
applicable) $(\vec \delta_k)_k=1$ and $(\vec \delta_k)_{k-1}=-1$. Each
tunnelling in the 'forward' direction is thus associated with a
phase factor $e^{i\phi/N}$. The corresponding supercurrent operator
is given by
\begin{equation}
I_{{\rm S},k}=\frac{(-2e)E_{{\rm J},k}}{2\hbar}\sum_{\vec n}(-i\vert
\vec n+\vec\delta_k\rangle\langle \vec n\vert e^{i\phi/N}+\,{\rm H.c.}\,)
\label{supercur}
\end{equation}
where $I_{{\rm c},k}\equiv(-2e)E_{{\rm J},k}/\hbar$ is the critical
current of junction $k$. We also define the
(average) supercurrent operator $I_{\rm S}$ by
\begin{equation}
I_{\rm S}=\frac1N\sum_{j=1}^NI_{{\rm S},j}=\frac{(-2e)}{\hbar}
\frac{\partial H}{\partial \phi}.\label{averagesup}
\end{equation}
The second equality follows from the $\phi$-independence
of $H_{\rm C}$ and the relation between $\phi_k$ and $\phi$.
The common expectation value of $I_{\rm S}$ and $I_{{\rm S},k}$
in a stationary state $\vert m\rangle$ is given
by $(-2e/\hbar)\partial E_m/\partial \phi$ where $E_m$ is
the corresponding eigenenergy. Since $I_{\rm S}$
can be expressed as a derivative of the full
Hamiltonian operator its matrix element between two different
stationary states $\vert m\rangle$ and $\vert l\rangle$
can be expressed simply as
\begin{equation}
\langle m\vert I_{\rm S}\vert l\rangle_\phi=\frac{(-2e)(E_m-E_l)}{\hbar}
\lim_{\phi'\rightarrow\phi}
\frac{{}_{\phi}\langle m\vert l\rangle_{\phi'}}{\phi'-\phi}.
\label{easycur}
\end{equation}
This expression is well-defined although one must keep track
of the physically unimportant total phases of the wave functions.

\subsection{The supercurrent and the transferred charge}

As shown in the previous article\cite{pek99} there are two mechanisms
of Cooper pair transfer in the array. The first one is the direct
supercurrent flowing through the whole array due to non-zero $\phi$
and the other one is pumping, the charge transfer in response to the
adiabatic variation of the injected charges $\vec q$.

The expressions for the charge transferred by these mechanisms can 
be derived as follows. For each instant of time $t$ we introduce the
basis of instantaneous eigenstates $\{\vert m_{(t)}\rangle\}$ with
eigenenergies $\{E_m\}$ of the full Hamiltonian
(\ref{simphami}) for a given $\vec q(t)$. Assuming
slowly varying  gate voltages we may solve
the time-dependent Schr\"odinger equation with the initial
condition $\vert \psi(t_0)\rangle=\vert m_{(t_0)}\rangle$
to obtain
\begin{eqnarray}
&&\vert\psi_{(t_0+\delta t)}\rangle=e^{-iE_m\delta t/\hbar}\vert 
m_{(t_0)}\rangle\cr
&&\ +\sum_{l(\ne m)}\frac{(e^{-iE_l\delta t/\hbar}-e^{-iE_m\delta t/\hbar})
\langle E_l\vert \vec\nabla_{\vec q}\rangle\cdot \frac{\partial \vec q}
{\partial t}
}{i(E_l-E_m)/\hbar}\vert l_{(t_0)}\rangle
\cr
&&\equiv\vert m_{(t_0)}\rangle+\vert \delta m_{(\delta t)}\rangle.
\end{eqnarray}
Here the term $\vert \vec\nabla_{\vec q}\rangle\cdot 
\frac{\partial \vec q}{\partial t}$ is the directional derivative
of the ground state with respect to the change in gate charges $\vec q$.
The amount of charge that passes through the junction $k$ during a
short time interval $\delta t$ is then
\begin{eqnarray}
\delta Q_k
&=&\int_{t_0}^{t_0+\delta t}
\langle \psi_{(t)}\vert I_{{\rm S},k}\vert\psi_{(t)}\rangle dt
=\delta t\langle I_{{\rm S},k}\rangle_{\vert m_{(t_0)}\rangle}\cr
&&+2{\rm Re}\left[\int_{t_0}^{t_0+\delta t}
\langle m_{(t_0)}\vert I_{{\rm S},k}\vert
\delta m_{(t-t_0)}\rangle dt\right],
\end{eqnarray}
where we have neglected the term quadratic in $\vert \delta m\rangle$
and oscillatory terms by assuming  that $\delta t\gg \hbar/(E_l-E_m)$
holds for all $l$.
The first term gives the charge transferred via direct supercurrent.
The second term, the induced charge transfer, can be integrated 
yielding
\begin{equation}
\delta Q_{k,{\rm ind}}=-2\hbar
\sum_{l(\ne m)}{\rm Im}\left[\frac{\langle m\vert I_{{\rm S},k}
\vert l\rangle\langle l\vert \delta m\rangle}{E_l-E_m}\right]
\end{equation}
where $\vert \delta m\rangle$ is the change in the instantaneous
eigenstate induced by the change $\vec q(t_0)\rightarrow \vec q(t_0+
\delta t)$.

For a closed path $\gamma$ the transferred charge must be equal for
all $N$ junctions so it can be written in terms of the average
supercurrent operator $I_{\rm S}$.  The total amount of charge, $Q$,
transferred over a pumping period $\tau$ is given by
\begin{eqnarray}
\frac Q{-2e}&=&\frac1{\hbar}\int_0^\tau 
\frac{\partial E_m(t)}{\partial \phi}dt\cr
&&-\frac{2\hbar}{-2e}\oint_{\gamma}
\sum_{l(\ne m)}{\rm Im}\left[\frac{\langle m\vert I_{{\rm S}}
\vert l\rangle\langle l\vert dm\rangle}{E_l-E_m}\right]
\label{numerpump}
\end{eqnarray}
where $\vert dm\rangle$ is the differential change of $\vert m\rangle$
due to a differential change of the gate voltages $d\vec q$. It should
be noted that the pumped charge depends only on the chosen path while
the amount of charge transferred by direct supercurrent also depends
on how the gate voltages are operated on the path. According to
Eq.~(\ref{easycur}) the pumped charge for the state $\vert m\rangle$
simplifies to
\begin{equation}
\frac{Q_{\rm p}}{-2e}=2\oint_{\gamma}
\sum_{l(\ne m)}{\rm Im}\left[\lim_{\phi'
\rightarrow\phi}\frac{{}_{\phi}
\langle m\vert l\rangle_{\phi'}}{\phi'-\phi}
\langle l\vert dm\rangle\right].\label{simppump}
\end{equation}
Thus the pumped current is mediated by the induced mixing of other
components into the initial state and modified depending on how the
relative phases of the eigenstates change with respect to the
diffential change in $\phi$. This formulation proves to be especially
effective in the regime of two-state dominance. The operating
frequency of gate voltages must satisfy $f\ll E_{\rm J}/\hbar$
so that the adiabatic approximation is valid, though.

\subsection{Numerical, renormalised and analytical results}

In the following sections we refer to our results as numerical,
renormalised or analytical. Numerical results are obtained by
diagonalising the Hamiltonian operator (\ref{simphami}) in a given
basis and using the corresponding eigenstates in order to evaluate the
required observable. The pumped charge $Q_{\rm p}$ is obtained by
numerically integrating the second term in Eq. (\ref{numerpump}).

Renormalised results are obtained by a semianalytical process where
the renormalised matrix elements in Eq. (\ref{vrenorm})
for a given $E$ are expressed
analytically  but the iteration process is
naturally done numerically.
Although restricting the analytical renormalisation into a given
basis is quite difficult it facilitates
direct comparison between the renormalised and
numerically obtained results. Purely numerical renormalisation
forfeits so much information that it is not used in this article.

The analytical results are obtained by renormalisation in
such a manner that they can be expressed in
a closed form. The analytical results for the supercurrent
are given by the relation $(-2e/\hbar)\partial E_m/\partial \phi$
while the pumped charge is evaluated using Eq. (\ref{simppump}).

\section{The homogeneous Cooper pair pump}\label{sec:homog}

\subsection{The properties of the charging Hamiltonian}\label{sec:charprop}

For uniform arrays all Josephson energies are
equal to $E_{\rm J}$ and $c_k\equiv1$ so the charging
energy is given by
\begin{equation}
E_{\vec n}=
E_{\rm C}\left[\sum_{k=1}^N{v_k^2}-\frac1N
\left(\sum_{k=1}^Nv_k\right)^2\right],
\label{unifchar}
\end{equation}
provided that $v_k$ are solutions of Eqs.~(\ref{veeesi}).
We shall now examine the properties and the symmetries
of the charging Hamiltonian $H_{\rm C}(\vec q)$.
Since the number of Cooper pairs on each island may obtain
only integer values the degree of symmetry of $H_{\rm c}(\vec q)$
depends on how exactly the gate voltages $\vec q$ match
the set $\{\vec n\}$.

The most symmetrical case is obtained when
$\vec q=\vec n_0$ for some
charge eigenstate $\vert \vec n_0\rangle$.
We then find $E_{\vec n_0}=0$ and
\begin{equation}
E_{\vec n_0\pm(\vec s)}=E_{\rm C}s(N-s)/N
\end{equation}
where $(\vec s)$ consists of
$s$ jumps through $s$ different junctions
in "forward" direction. Thus the Cooper pair pump is single-state
dominant near the point of high symmetry and the
supercurrent is given by $(-2e/\hbar)\partial\tilde V_{11}/\partial\phi$.
The ``average-0'' choice now yields a supercurrent
$I_{\rm C}\cos(\phi)(N\varepsilon_{\rm J}/2)^{N-1}N/(N-1)!$
in agreement with numerical results at $\vec q=\vec n_0$.

The degree of symmetry is almost as high on any line $\vec
n_0+x\cdot\vec\delta_r$ for any $r$ and $x\in[0,1]$.  All junctions
except the junction $r$ are equivalent with respect to the charging
energy. In the region $x\approx 0.5$ the system is manifestly
two-state dominant. This limit is the most relevant one for us since
it is realised by ideal saw-tooth gating shown in Fig.~\ref{fig:cppump}.

The Hamiltonian is symmetric also at the so-called resonance
point $\vec q=\vec n_0+(1/N,\ldots,1/N)$, where the states
$\{\vert\vec n_0+\sum_{j=1}^{k}\vec\delta_j\rangle\}_{k=1}^{N}$ become
degenerate. For $N$ degenerate levels
with nearest-neighbour coupling the ground state supercurrent
is given by  $I_{\rm res}^{(0)}(\phi)\equiv I_{\rm C}\sin(\phi/N)/N$
with $\phi\in(-\pi,\pi)$ exhibiting a cusp at $\phi=\pm\pi$.
Because $\varepsilon_{\rm J}>0$
the coupling to the other states enhances the supercurrent
which will be explicitly evaluated in Sec.~\ref{sec:bases}.

\subsection{Pumping and supercurrent for homogeneous arrays}

We will now evaluate the inaccuracy in the pumping for the
uniform array when the gate voltages are operated as
depicted in Fig.~\ref{fig:cppump}. Due to the symmetry of
the charging Hamiltonian $H_{\rm C}$ it is
enough to consider any one of the legs and multiply
the results by $N$. In the Coulomb blockade regime for the
saw-tooth gating cycle the system is always
dominated by either one or two charge eigenstates.
The pumping mainly occurs when these two states for each leg
are nearly degenerate.

A two-level Hamiltonian can always be decomposed as
\begin{equation}
H=\left(\matrix{\epsilon_1& ve^{-i\theta(\phi)}\cr ve^{i\theta(\phi)}
&\epsilon_2}\right).
\end{equation}
For the truncated two-level system we have
$\theta(\phi)=\phi/N$, $v=-E_{\rm J}/2$ and $\epsilon_j=E_{\rm ch}^{(j)}$,
$j=1,2$. The proper decomposition of the  renormalised Hamiltonian is
\begin{equation}
\tilde H\approx\left(\matrix{E_{\rm ch}^{(1)}-a^{(1)}
\cdot E_{\rm C}&
v\vert b(\phi)\vert e^{-i(\phi/N+\phi_b)}\cr
v\vert b(\phi)\vert e^{i(\phi/N+\phi_b)}&
E_{\rm ch}^{(2)}-a^{(2)}\cdot E_{\rm C}}\right)
\label{tworeno}
\end{equation}
where $v=-E_{\rm J}/2$, $a^{(j)}=a^{(j)}_0+a^{(j)}_1\cos\phi$,
$j=1,2$, $b=b_0+b_{-1}e^{-i\phi}+b_1e^{i\phi}$ and $e^{i\phi_b}=
b/\vert b\vert$. The leading
components for these coefficients  are  $a_0\propto\varepsilon_{\rm J}^2$,
$a_1\propto\varepsilon_{\rm J}^N$, $b_0\approx 1+c\varepsilon_{\rm J}^2$,
$b_{-1}\propto\varepsilon_{\rm J}^{N-2}$ and $b_1\propto
\varepsilon_{\rm J}^N$. Thus $\tilde H$ clearly tends to the
unrenormalised Hamiltonian $H$ in the 
natural limit $\varepsilon_{\rm J}\rightarrow0$.

The actual values for these parameters are discussed below. The next 
corrections in $\tilde H$ are $a_{2}$ and $b_{\mp2}$ which are further
suppressed by a factor $\varepsilon_{\rm J}^N$. The decomposition
(\ref{tworeno}) is valid also for inhomogeneous Cooper pair pumps
with only superficial changes. The renormalisation process itself
is more complicated, though.  

In the context of this two-level model the renormalisation coefficients
$a$ and $b$ have the following interpretation. The diagonal
coefficient $a_0$ corresponds to those tunnelling sequences
that end in the same state inside the active $P$-space and
without transporting any Cooper pairs through the array.
Coefficients $a_1$ correspond to those sequences that
transport one Cooper pair in forward or backward direction,
thus yielding the $\cos(\phi)$-dependent term. The coefficients
$b$ arise from the sequences that connect the two
charge eigenstates via the $Q$-space, transporting
$-1$, $0$ or $1$ Cooper pairs through the array.
Each intermediate state $\vert \vec n\rangle$ naturally
introduces an energy denominator $E-E_{\vec n}$.

This interpretation is illuminating and extremely helpful when
evaluating the renormalisation  coefficients but it has a
very severe drawback. The picture we obtain is, unfortunately,
false. The quantum mechanics implies that if $\varepsilon_J\ne0$ all
charge eigenstates simultaneously coexist although the
amplitude of most of these states is negligible in the low-lying
eigenstates of the system. 
The renormalising sequences appear when we take into
account the existence of high-lying states as described in
Sec.~\ref{sec:effect}.

In order to calculate the integral (\ref{simppump})
for the leg in pumping we will use  a parameter
$\eta=(\epsilon_1-\epsilon_2)/2v$ which is linear in
the ascending gate voltage for the truncated system and
almost linear for the renormalised system. The
gating induced correction for the ground state
yields a term $\langle 2\vert d1\rangle=\frac12d\eta/(1+\eta^2)$
which is real. Thus we only need the imaginary part
of the limit in (\ref{simppump}) which reads
\begin{equation}
{\rm Im}\left[\lim_{d\phi\rightarrow0}\frac{{}_{\phi_0}
\langle 1\vert 2\rangle_{\phi_0+d\phi}}{d\phi}\right]=
-\frac{d\theta/d\phi}{2\sqrt{1+\eta^2}}.
\end{equation}
For the truncated system $d\theta/d\phi=1/N$ and the pumped
charge is given by
\begin{equation}
\frac{Q_{\rm p}}{-2e}=\frac1{2N}\left[\frac{\eta_i}{\sqrt{1+\eta_i^2}}-
\frac{\eta_f}{\sqrt{1+\eta_f^2}}\right].\label{truncpump}
\end{equation}
The symmetry of the full Hamiltonian (\ref{simphami}) implies that the
pumped charge for the full cycle should be exactly $-2e$ so we bluntly
assume that the charge transfer in the limit $\theta\rightarrow
\phi/N$ is exactly $Q_{\rm p}=-2e$. We can partially justify this
assumption by allowing for the missing charge transfer via higher
excited states and noting that the identification of the initial and
final states changes after each leg.

For the renormalised system $d\theta/d\phi$ may be evaluated
analytically yielding
\begin{equation}
\frac{d\theta_{\rm ren}}{d\phi}=\frac1N+\frac{b_0(-b_{-1}+b_{1})
\cos\phi-b_{-1}^2
+b_1^2}{\vert b(\phi)\vert^2}.\label{renopump}
\end{equation}
The pumping inaccuracy is then given by a weighted average of
$Nd\theta/d\phi$ on a single leg. The weights can be obtained
from Eq.~(\ref{truncpump}) but for practical purposes it
suffices to evaluate Eq. (\ref{renopump}) at the degeneracy point.
The coefficients are obtained by using the
``average'' choice for the eigenenergy $E$.
In most cases the renormalisation includes all
terms up to the third order and coefficients $a_1$ and
$b_{\mp1}$ up to and including order $\varepsilon_{\rm J}^N$.
The leading correction from $b_{-1}\cos\phi$ is proportional to
$\varepsilon_{\rm J}^{N-2}$ as shown in Ref. 4.

In Fig.~\ref{fig:n3inacc} the pumped charge $Q_{\rm p}$ for $N=3$ is
studied as function of the phase difference $\phi$. The renormalised and
numerical results are in good agreement and they clearly indicate that
the deviations from the leading order result ($Q_{\rm
p}/(-2e)=1-9\varepsilon_{\rm J}\cos\phi$) are important even when
$\varepsilon_{\rm J}$ is relatively small.
The maximum value for the numerical and
renormalised pumped charges are $2.14$ and $2.18$, $2.87$ and $2.98$,
$3.70$ and $3.89$ for $\varepsilon_{\rm J}=0.1$, $0.15$ and
$0.2$, respectively.

Both minimum and maximum values of $Q_{\rm p}$ correspond to vanishing
supercurrent which suggests that, in principle, the phase differences
$\phi=0$ and $\phi=\pi$ can be differentiated and the ratio $Q_{\rm
p,max}/Q_{\rm p,min}$ could be used in order to determine
$\varepsilon_{\rm J}$. More realistic models are required
in order find out if this signature can persist when
effects due to the electromagnetic environment are included.

In units $E_{\rm C}$
the renormalised eigenenergies read
\begin{equation}
\left.\matrix{\tilde E_{1}\cr
\tilde E_{2}}\right\}=\frac{\epsilon_1+\epsilon_2}2\mp
\hbox{$\frac12$}\sqrt{(\Delta\epsilon)^2+\varepsilon_{\rm J}^2
\vert b(\phi)\vert^2}
\end{equation}
where $\Delta\epsilon=\epsilon_1-\epsilon_2$.
The ground state supercurrent is obtained by deriving
$\tilde E_1$ with respect to $\phi$ with result
\begin{eqnarray}
&&\langle I_{\rm S}\rangle_{\rm g.s.}
=(I_{c}\sin\phi)\left[\frac{a^{(1)}_1+a^{(2)}_1}{2
\varepsilon_{\rm J}}\right.\cr
&&+\left.\hskip-2pt\frac{b_0(b_{-1}+b_1)\hskip-1.5pt+
\hskip-1.5pt2b_{-1}b_1\cos\phi\hskip-1.5pt-\hskip-1.5pt
(a^{(1)}_1-a^{(2)}_1)\Delta\epsilon/\varepsilon_{\rm J}}
{(2/\varepsilon_{\rm J})\sqrt{(\Delta\epsilon)^2+\varepsilon_{\rm J}^2
\vert b(\phi)\vert^2}}\right].\label{gensupercur}
\end{eqnarray}
In Fig.~\ref{fig:supercur} the renormalised and numerical
supercurrents are shown for $N=5$. The renormalisation using the
``individual'' choice reproduces the supercurrent well in all three
cases. For clarity a basis with 40 states was used although it is not
large enough to produce the leading order ($\varepsilon_{\rm
J}^{N-1}$) supercurrent fully.
The slight underestimation of the supercurrent at the
degeneracy point is explainable since we could use only
leading order terms in the renormalisation.
Below we study various effects related to restricted bases.

\subsection{The basis-dependent effects}\label{sec:bases}

In order to reliably evaluate the pumped charge or
the supercurrent one must first select a proper basis in
which the calculations are performed.
The basis should be as large as possible so that
the discarded states are not important but on the
other hand such calculations may require prohibitive
amounts of CPU-time. Our aim is to circumvent these problems
by using renormalisation techniques.

Numerical calculations have been mainly performed in three
classes of bases we refer to as a-, b- and c-bases.
From here on we reserve the superscript in parenthesis
$^{(k)}$ for the $\varepsilon_{\rm J}^k$-dependent part of any
coefficient and the distinction between charge eigenstates $1$ and
$2$ is taken to be complied implicitly. An a-basis contains
all states contributing to the leading order inaccuracy
($b_{-1}^{(N-2)}$) while a b-basis produces the leading component
of the supercurrent ($a_{1}^{(N)}$) fully. Even larger c-basis
contains all states required for the next-to-leading correction
of the inaccuracy ($b_{-1}^{(N)}$, $b_{1}^{(N)}$).

For each length of the array $N$ these bases can be created as
follows. The leading component of the inaccuracy (supercurrent) is
carried by the $N-1$-step ($N$-step) ``paths'' containing at most one
tunnelling through each junction that connect the initial state to the
final state (itself) for each leg. The total number of necessary
states is $2^{N-2}N$ and $2^{N-1}N$ for an a-basis and a b-basis,
respectively. A short reasoning confirms that a state should be
included in a c-basis if it can be reached from some state in the
corresponding b-basis by a single tunnelling. The number of states has
not been generally resolved but in cases $N=4$ to $N=8$ the c-basis
contains $100$, $325$, $966$, $2695$ and $7176$ states.

The numerical 
evaluation of the pumping integral in Eq.~(\ref{numerpump}) may
require up to $10^5$ to $10^6$ complete diagonalisations per leg
before convergence is obtained which is extremely time-consuming for
bases containing more than few hundred basis states.  The supercurrent
has been evaluated even for the $N=8$ c-basis since only one
eigenstate is needed. Due to computational necessities some
modifications of the above-mentioned bases have been used.

The differences between bases can be illuminated by performing 
an ``average-0'' choice renormalisation at the degeneracy point of 
the saw-tooth gating path. Inserting the coefficients
$b_{j}^{(k)}$ ($j=0,\pm1$) and respective powers of 
$\varepsilon_{\rm J}$ into Eq.~(\ref{renopump}) one obtains
a power expansion of the inaccuracy for small  
values of $\varepsilon_{\rm J}$. This expansion has to be corrected
for the drop in ground state energy induced by the terms $a_0^{(k)}$.

In order to include all contributions up to the next-to-leading order
$\varepsilon_{\rm J}^{N}$ we need the expressions for coefficients
$a_0^{(2)}$, $b_0^{(2)}$, $b_{-1}^{(N-2)}$, $b_{-1}^{(N)}$ and
$b_1^{(N)}$. Simple expressions are obtained for
\begin{eqnarray}
a_{0,{\rm a}}^{(2)}&=&\frac{N-2}4+\frac{N(N-1)}{4(N-2)},\cr
a_{0,{\rm b}+}^{(2)}&=&\frac{N-1}4+\frac{N(N-1)}{4(N-2)}+
\frac{N}{4(2N-2)},\cr
b_{0,{\rm a}}^{(2)}&=&N/2,\cr
b_{0,{\rm b}+}^{(2)}&=&N(N-1)/2(N-2),\cr
b_{-1,{\rm a}+}^{(N-2)}&=&\left(\frac N2\right)^{N-2}\frac{N-1}
{(N-2)!}\nonumber
\end{eqnarray}
where index a and b corresponds to a- and b-bases and $+$ implies that
coefficient does not change when basis is enlarged.  The analytical
expressions for the coefficients $b_{-1}^{(N)}$ and $b_{1}^{(N)}$ are
composed of several multiple summations.  The obtained 
values of $b_{-1}^{(N)}$ for different bases and $b_{1}^{(N)}$ for c-basis
are given in Table~\ref{tab:coeffs}.

The power expansion of the pumped charge then reads
\begin{eqnarray}
&&\frac{Q_{\rm p}}{-2e}\approx1-N\varepsilon_{\rm J}^{N-2}
\cos\phi\left[b_{-1}^{(N-2)}
\right.\cr
&&\ +\varepsilon_{\rm J}^2\left.\left(
b_{-1}^{(N)}-b_{1}^{(N)}-(Nl(N)a_0^{(2)}+b_0^{(2)})b_{-1}^{(N-2)}
\right)\right]
\label{pumpapprox}
\end{eqnarray}
where $l(N)\approx1$ stems from the energy denominators. Its value is
$1$, $1$, $11/12$, $5/6$ and $137/180$ in cases $N=3$ to $N=7$,
respectively. For $N=3$ and $N=4$ the strong deviations from
$\cos(\phi)$-dependence are explained by additional terms
$27(\varepsilon_{\rm J}\cos\phi)^2-81(\varepsilon_{\rm J}\cos\phi)^3$
and $24\varepsilon_{\rm J}^4\cos^2\phi$, respectively.  The expansion
(\ref{pumpapprox}) for $N=3$ does not compare too well against
numerical results in Fig.~\ref{fig:n3inacc} but inclusion of the
above-mentioned terms improves agreement considerably up to
$\varepsilon_{\rm J}\approx0.1$.

Next we take a closer look at the case $N=5$ in Fig.~\ref{fig:n5inacc}
where the power expansions for the b-basis and the $240$-state basis
(almost full c-basis\cite{note1}) as well as the results for ``average''
renormalisation are compared to numerical results for $\phi=0$. 
The inaccuracy is given in units $\varepsilon_{\rm J}^3$ which
allows more detailed comparison of the predictions. The renormalised
values follow the numerical results more closely than the power
expansions but the differences between bases are still reproduced
well up to $\varepsilon_{\rm J}\approx0.1$. In addition the 
inaccuracy for a-basis is correctly placed in between these bases.

Similar overestimation can be seen in the inset of
Fig.~\ref{fig:n5inacc} showing the numerical and renormalised
inaccuracies for $N=7$.  Although the results may not seem to be so
good at the first glance, one should bear in mind that the 336-states
basis for which the convergence is the best corresponds to even
smaller an inaccuracy than the $N=7$ b-basis. Actual inaccuracy should
be evaluated for much larger c-basis which is, unfortunately, clearly
impossible. The scaling of the inaccuracy by $\varepsilon_{\rm J}^5$
certainly exaggerates the error, too.  In conclusion we may state that
the renormalisation seems to be able to reproduce the behaviour of the
inaccuracy reasonably well for any $N$ and $\varepsilon_{\rm J}$ in
the Coulomb blockade regime.

The enhancement of the supercurrent at the resonance
point has been studied but the conclusions
remain valid also in its vicinity. For small values
of $\varepsilon_{\rm J}$ next-to-nearest neighbour
coupling yields approximate supercurrent
$1+N\varepsilon_{\rm J}\cos(\phi/N)$ in our units
of choice, $I_{\rm res}^{(0)}(\phi)$.
In the more general case have used semianalytical third order
renormalisation with $2N(N-1)$ state $P$-space and compared the
results to the supercurrent obtained by diagonalisation.

For $\phi\approx\pi$ and $6\le N\le 10$ and  the comparison is shown
in Table~\ref{tab:renor} clearly indicating that the differences
between bases for $\varepsilon_{\rm J}=0.1$ are not significant but for
$\varepsilon_{\rm J}=0.2$ they are growing. The renormalisation
calculations indicate that for $\varepsilon_{\rm J}=0.1$ the
convergence is fast both with respect to the order of renormalisation
as well as the basis.
As conclusion we may state that the enhancement is
important for large $N$ and $\varepsilon_{\rm J}$ although it will not
cancel the overall suppression $\sim 1/N^2$ of the maximal
supercurrent.

\section{Inhomomogeneity in the array}\label{sec:inhomog}

In this section we will derive the leading contributions for the
pumping inaccuracy and the supercurrent on the saw-tooth gating path
for an inhomogeneous array.  General considerations imply that the
quantity $E_{\rm J}E_{\rm C}$ is approximately constant for all
junctions in an array. The Josephson energy is inversely proportional
to the normal state resistance $R_{\rm T}$ of the junction and $E_{\rm
C}$ is inversely proportional to the capacitance $C$ of the
junction. Since $R_{\rm T}$ is inversely proportional and $C$ is
directly proportional to the area of the junction, the product is
approximately constant for different junctions in the array. (This
argument works for junctions fabricated in the same batch;
otherwise the constants of proportionality are different.)

The model Hamiltonian (\ref{simphami}) is uniquely defined by the
ratio $\varepsilon_{\rm J}$ and relative capacitances
$\{c_k\}_{k=1}^N$ once we set $E_{{\rm J},k}=c_kE_{\rm J}$ in the
tunnelling Hamiltonian $H_{\rm J}$. We define
the inhomogeneity index of the array
\begin{equation}
X_{\rm inh}=\left(\frac1N\sum_{k=1}^N g_k^2\right)^{1/2}
\label{inhomogene}
\end{equation}
where $g_k=1/c_k-1$ in order to
study the behaviour of the inaccuracy as a function of $X_{\rm
inh}$. Although the definition is valid for arbitrary $X_{\rm inh}$
the limits $X_{\rm inh}<0.15$ and $\vert g_k\vert<0.5$
are reasonable for the current technology at capacitances of the order
of 1 fF.

On $r^{\rm th}$ leg of the
gating cycle the initial and final states are
$\vert\vec n_r\rangle$ and $\vert\vec n_r+\vec\delta_r\rangle(
\vert\vec n_{r+1}\rangle)$, respectively.
The gate voltages are given by $\vec q=\vec n_r+x\vec \delta_r$
where  $x\in[0,1]$ is the normalised ascending gate voltage so that
$x=\frac12$ corresponds to the degeneracy point.
In order to obtain the leading order contributions for
the inaccuracy and supercurrent we must set $a_0=0$, $b_0=c_r$
in the renormalised Hamiltonian (\ref{tworeno}) and
evaluate coefficients $a_{1}$ and $b_{-1}$.

All the required states on the $r^{\rm th}$ leg can be
chosen from the classes $\vert\vec n_r\pm(\vec s)\rangle$ for $s=0,1,
\ldots,N-1$ where $(\vec s)$ defined in Sec.~\ref{sec:charprop} does
not contain $\vec\delta_r$. Let $\sigma$ denote a permutation of the
set $\{1,2,\ldots,N\}\backslash\{r\}$, $\sigma(s)$ the set of $s$
first elements in $\sigma$ and $\sigma_k$ the $k^{\rm th}$ element of
$\sigma$.\cite{note2} Each $\sigma(s)$ then
defines two states with charging energies
\begin{eqnarray}
E_{\pm s,x}&\equiv&E_{\vec n_r\pm(\vec s)}(x,\sigma)=\frac{E_{\rm C}}N\left[
\left(N/c_r-1/c_r^2\right)x^2\right.\cr
&&\left.+(s+G_{\sigma}^s)\left((N-s)\pm 2x(1+g_r)-G_{\sigma}^s
\right)\right]
\label{simpcharge}
\end{eqnarray}
where $G_{\sigma}^s=\sum_{k=1}^sg_{\sigma_k}$. The charging energies 
for the initial and final states are given by
$E_{0,x}$ and $E_{0,1-x}$,
respectively.

The leading order of pumped charge can be obtained for the ``average-0''
choice at the degeneracy point for each leg yielding result
\begin{eqnarray}
\frac{Q_{\rm p}}{-2e}&=&1-(K\cos\phi)
\sum_{r=1}^N\sum_{\sigma}
\frac{c_r^{-2}}{\prod_{s=1}^{N-2}\Delta E_{-s,\sigma}},
\label{inhinaccu}
\end{eqnarray}
exact in the limit $\varepsilon_{\rm J}\rightarrow0$.
Here $K\equiv \left(\frac{NE_{\rm J}}
{2E_{\rm C}}\right)^{N-2}\prod_{k=1}^Nc_k$ and
\begin{eqnarray}
\Delta E_{-s,\sigma}&\equiv&(N/E_{\rm C})
(E_{-s,\frac12}-E_{0,\frac12})\cr
&=&(s+G_{\sigma}^s)(N-s-1-g_r-G_{\sigma}^s).
\end{eqnarray}
The analytical result~(\ref{inhinaccu}) gives us the theoretical ratio
between inhomogeneous and homogeneous inaccuracy which will be denoted
by $W_{\rm inh}$ and compared to numerical results.  The
interpretation is obvious since for small values of $\varepsilon_{\rm
J}$ the higher order corrections are not very important and even then
their behaviour is relatively similar to the leading contribution.

The inaccuracy is invariant under arbitrary permutations of the set
$\{c_k\}$. Numerically the pumped charge may be evaluated for any of
the $N$ junctions but far better numerical convergence is obtained by
using the average supercurrent operator $I_{\rm S}$. The total
inaccuracy for an inhomogeneous array is always larger than the
corresponding homogeneous array and the junctionwise inaccuracy for
$c_k>1$ ($c_k<1$) is smaller (larger) than average inaccuracy.

In Fig.~\ref{fig:inhinaccu} $W_{\rm inh}$  is plotted as
function of $X_{\rm inh}$ corresponding to some specific
sets of relative capacitances $\{c_k\}$ for $N=4$ and $N=5$.
The numerical results have been obtained for b-bases.
The agreement between analytical and numerical results
is good showing that the effects due to inhomogeneity of the array
can be reliably treated as a correction factor when relative
capacitances $c_k$ are given.

The effects due to inhomogeneity can be parametrised by
obtaining limits for $W_{\rm inh}$ as a function of
$X_{\rm inh}$. For $X_{\rm inh}=\vert g\vert$ this
is achieved by considering the even distribution of inhomogeneity
($g_{\rm odd}=g$, $g_{\rm even}=-g$ for even $N$ and
$g_{\rm odd}=g\left(\frac{N-1}{N+1}\right)^{1/2}$,
$g_{\rm even}=g\left(\frac{N+1}{N-1}\right)^{1/2}$
for odd $N$) yielding a lower limit and maximally
distorted distribution ($g_1=g\tilde N$,
$g_{k(\ge2)}=g/\tilde N$, $\tilde N=\sqrt{N-1}$)
corresponding to an upper limit. The upper limit
yields a simple, analytical result
\begin{equation}
W_{\rm inh}(X_{\rm inh},N)\le\max[f(X_{\rm inh},N),
f(-X_{\rm inh},N)]
\label{eq:inhlimit}
\end{equation}
where
\begin{eqnarray}
&&f(g,N)=\frac{[1-g/\tilde N]^{5-3N}}{N(1+g\tilde N)}
\left[(1+g\tilde N)^2\right.\cr
&&\left.\ +(1-g/\tilde N)^2
\sum_{k=1}^{N-1}\left[\prod_{s=N-k}^{N-2}
\frac{s}{s+\gamma}\prod_{s=k}^{N-2}
\frac{s}{s+\gamma}
\right]\right]\label{eq:gfun}
\end{eqnarray}
with $\gamma=Ng/(\tilde N-g)$. The analytical
expression for the lower limit is obtained by explicitly
inserting the even distribution in Eq.~(\ref{inhinaccu})
and using the symmetry in order to reduce the number of
terms to be calculated. Even simpler a lower limit can
be obtained by considering the asymptotical behaviour of the
inhomogeneity. We find
\begin{equation}
W_{\rm inh}(X_{\rm inh},N)\ge1+a_N^{({\rm inh})}X_{\rm inh}^2
\label{eq:asymplim}
\end{equation}
where the $N$-dependent constant $a_N^{({\rm inh})}$
can be evaluated from Eq.~(\ref{eq:gfun}) yielding
values $8$, $85/9$, $1279/20$, $42317/3600$, $40267/3150$,
and $13.769$ for cases $N=4$ to $N=9$, respectively.\cite{note4}
In Fig.~\ref{fig:inhlimits} we graphically present the
limits for  $(W_{\rm inh}-1)/X_{\rm inh}^2$ as function
of $X_{\rm inh}$ in cases $N=4$ to $N=7$.
For $X_{\rm inh}=0.15$ we obtain
limits $20$ \%, $24$ \%, $28$ \% and $32$ \% as the
maximal increase in inaccuracy as compared to the
homogeneous case for $N=4$ to $N=7$, respectively.

The leading order renormalised supercurrent may be
evaluated using the "individual-0" choice as follows. 
The eigenenergies of the truncated Hamiltonian are
\begin{equation}
\left.\matrix{\tilde E_1\cr \tilde E_2}\right\}
=\frac{E_{0,x}+E_{0,1-x}}2\mp\hbox{$\frac12$}\sqrt
{\Delta^2_{0,x}+c_r^2E_{\rm J}^2}
\end{equation}
where $\Delta_{0,x}=E_{0,x}-E_{0,1-x}$. 
Using $\tilde E_1$ in the renormalisation now yields
the leading order supercurrent $I_{r,x}\equiv
\langle I_{\rm S}\rangle_{(r,x)}$, on leg $r$ for
ascending gate voltage $x$ as
\begin{equation}
I_{r,x}=\sum_{\sigma}\sum_{l=1}^{N}\frac{(I_c\sin\phi)\left(E_{\rm J}/
2\right)^{N-1}\prod_{k=1}^Nc_k}
{\left(\prod_{m=1}^{l-1}\Delta E_{\sigma}^{(m)}
\right)\left(\prod_{m=l}^{N-1}
\Delta E_{\sigma}^{(r,m)}\right)}
\label{inhosuppi}
\end{equation}
where energy differences are given by
$\Delta E_{\sigma}^{(s)}=E_{s,x}-\tilde E_1$,
for $s=1,\ldots,N-1$ and $\Delta E_{\sigma}^{(r,s)}=
E_{\bar s,x}-\tilde E_1$, $\bar s=-(N-s-1)$ corresponding
to the last elements of $\sigma$.
From Eq.~(\ref{gensupercur}) we find $\Delta E_{\sigma}^{(r,0)}=
\tilde E_2-\tilde E_1$ for the remaining
energy difference.
In Fig.~\ref{fig:inhosup} the analytical prediction (\ref{inhosuppi})
is compared against and numerically evaluated  supercurrent
for $N=6$ b-basis. Each curve corresponds to
a randomly chosen set $\{c_k\}$ $\varepsilon_{\rm J}\in[0.02,0.06]$
as seen from the different
widths of the peak. The numerical and
analytical results practically coincide.

Finally, by using (\ref{inhosuppi}) we can explain the
$a$-basis supercurrent in Fig.~\ref{fig:supercur}.
For a homogeneous array $g_k\equiv0$ and the
charging energies $E_{\pm s,x}$ are identical so the summation
over $\sigma$ yields a prefactor $(N-1)!$. The omission of
certain states amounts to disallowing some paths
and the numerator has to be corrected by a factor $(l-1)/(N-1)$
which is exactly how the supercurrent for $a$-basis
can be evaluated.

\section{The pumping inaccuracy and nonideal gating}\label{sec:practic}

In this section we will derive the leading order
correction induced by nonideal gating sequences
which we  define below. When all gate voltages are
turned off in Fig.~\ref{fig:cppump} in beginning of
the first leg the actual gate voltages can be expressed
as $\vec q_{(1)}=\vec n_1+\vec q_{\rm off}$ defining
the $N-1$ offset errors. The maximum values of the
normalised sweeping voltages
are given by $q_{{\rm sw},k}=1+\tilde q_{\rm sw}^{(k)}$ whence
the initial gate voltages on $k^{\rm th}$ leg read
\begin{equation}
\vec q_{(k>1)}=\vec n_k+\vec q_{\rm off}+\tilde q_{\rm sw}^{(k-1)}
\vec\Delta_{k-1}
\end{equation}
where $\vec\Delta_{k}=\sum_{j=1}^k\vec \delta_k$.
In general, the sweeping voltages can be determined more
precisely than the offset voltages, and for most of
our calculations we have used 1 \% and 2 \% precisions
for them, respectively. 

As in the case of inhomogeneity we can determine the effects
due to nonideal gating sequences provided we can 
evaluate the charging energy differences at the
degeneracy point. On the $r^{\rm th}$ 
leg we choose the coordinates of the degeneracy point as 
\begin{equation}
\vec q_{(\rm deg,r)}=\vec n_r+\hbox{$\frac12$}\vec\delta_r+
\sum_{j=1}^N\mu_j\vec\delta_j
\end{equation}
thus defining quantities $\mu_j$ subject to condition $\mu_r=0$. 
The degeneracy condition $E_{\vec n_r}=E_{\vec n_{r+1}}$ can be
solved easily yielding
\begin{equation}
\sum_{j=1}^N(\mu_j/c_j)=0.\label{eq:hyperplane}
\end{equation}
Thus on each leg one must find where the line connecting the initial
and final gate voltages crosses the hyperplane defined
by Eq.~(\ref{eq:hyperplane}). This clearly implies
that the  correct nonideality parameter is
\begin{equation}
X_{\rm non}=\left[\frac1N\sum_{{\rm leg}=1}^N\sum_{j=1}^N
\left(\frac{\mu^{({\rm leg})}_j}{c_j}\right)^2\right]^{1/2}
\end{equation}
which can easily be evaluated once the offset and sweeping voltages
are given.

The leading order inaccuracy may be evaluated using
the charging energies (\ref{generchar0}) at the degeneracy
point and inserting the corresponding energy differences
(multiplied by $N/E_{\rm C}$) into Eq.~(\ref{inhinaccu}).
Dividing the result by the inaccuracy for a homogeneous
array and ideal gating sequence we obtain the ratio
$W_{\rm non}$. We are mainly interested in the behaviour
of $W_{\rm non}$ for very small values of $X_{\rm non}$
so we have only evaluated the asymptotical limit
\begin{equation}
W_{\rm non}\sim 1+a_N^{({\rm non})}X_{\rm non}^2
\label{eq:nonidlim}
\end{equation}
for homogeneous array. Here  $a_N^{({\rm non})}$ is given by
$40/3$, $1225/108$, $41/4$, $258181/27000$, $6136/675$ in cases $N=4$
to $N=9$, respectively.

In order to show that $X_{\rm non}$ really is the
correct parameter we chose several sets of offset and
sweeping voltages for homogeneous arrays. We then
evaluated $W_{\rm non}$
both analytically and numerically for $N=4$ a-basis
with $\varepsilon_{\rm J}=0.05$ and
$N=5$ a-basis with $\varepsilon_{\rm J}=0.04$. The results
are shown in Fig.~\ref{fig:nonideal} which also show
the asymptotic limits. Theoretical values lie on the
curves as well as most of the numerical data points.

The full inaccuracy may be approximately
understood in terms of the contribution
from the inhomogeneity and nonideality. In order to show this
we have used nonideal gating sequences with some of the
inhomogeneous arrays already used in Fig.~\ref{fig:inhinaccu}.
We have collected the results in Table~\ref{tab:inhonon} which includes
the parameters $X_{\rm inh}$ and $X_{\rm non}$ and
numerical ratio $W_{\rm non}$ as compared to the
full theoretical correction and product of corrections due
to pure inhomogeneity and pure nonideality.
The reasonable agreement between theoretical
and numerical results shows that we really can
take into account both inhomogeneity and nonideal gating
sequences.

Finally we must note that since the connection between $X_{\rm non}$
and offset and sweeping voltages is much more complicated than the
connection between relative capacitances and $X_{\rm inh}$, there is
no straighforward way to obtain $X_{\rm non}$ from the
experimental data. An approximate upper limit can be given easily,
though. A sort of worst-case scenario for $\vert q_{\rm off,k}+\tilde 
q_{\rm sw}^{(k)}\vert <x_{\rm non}$ when the precision
of the gating for each component is known, yields
\begin{equation}
X_{\rm non}(N)\approx \frac{x_{\rm non}(14+11N+4N^2+N^3)^{1/2}}{6^{1/2}N}
\end{equation}
which can be used in Eq.~(\ref{eq:nonidlim}) and combining
this result with the estimated limits for $X_{\rm inh}$ one
obtains reasonable limits for the ratio $W_{\rm non}$. Multiplying
$W_{\rm non}$ by the inaccuracy corresponding to homogeneous
array and ideal gating, allowing for indeterminacy of $\varepsilon_{\rm J}$, 
finally yields the final prediction
of the present model. The prediction, based on these three parameters, is a 
range inside which the inaccuracy is expected to lie, but 
it remains to be seen whether the electromagnetic environment 
or other effects strongly modify the present results.

\section{Conclusions}\label{sec:conclu}

We have studied pumping of Cooper pairs for an unbiased array
of Josephson junctions in an environment with vanishing
impedance. The present model,
which includes only charging effects and Cooper pair tunnelling,
can be reliably solved yielding relatively simple predictions
for the direct supercurrent and the accuracy of the pumping of 
Cooper pairs. 

We have successfully evaluated higher order corrections
for the supercurrent as well as the pumping inaccuracy
for ideal gating sequence and homogeneous arrays.
The effects due to inhomogeneous arrays or nonideal
gating sequences can be quantitatively treated by
defining parameters $X_{\rm inh}$ and $X_{\rm non}$ 
and respective correction factors $W_{\rm inh}$ and $W_{\rm non}$.

The parameters $\varepsilon_{\rm J}$ and $X_{\rm inh}$ can
be experimentally measured and the precision of the 
gate voltages yields limits for $X_{\rm non}$ so the
present model can give an explicit prediction for the
expected range of the experimental inaccuracy. The theoretical 
predictions have been verified by numerical calculations, but whether 
the model is realistic enough to give quantitatively,
or least qualitatively correct results will be ultimately
tested in experiments. In any case, further theoretical
studies using more realistic and sophisticated models 
should be performed.

\vbox{
\begin{table}[hbt]
\caption{The oefficients
$b_{-1}^{(N)}$ for a-, b- and c-basis and
$b_1^{(N)}$ for c-basis in cases $N=3$ to $N=9$ obtained by
using ``average-0'' choice renormalisation at the degeneracy
point. Exact values are given
as fractions when the value fits into the column.
\label{tab:coeffs}}
\vspace{0.2cm}
\begin{center}
\footnotesize
\begin{tabular}{ccccc}
{\raisebox{0pt}[5pt][5pt]{$N$}}&$b_{-1,{\rm a}}^{(N)}$&
$b_{-1,{\rm b}}^{(N)}$&$b_{-1,{\rm c}}^{(N)}$&
$b_{1,{\rm c}}^{(N)}$\cr
\hline
\raisebox{0pt}[7pt][0pt]{3}&$9/4$&$57/8$&$69/8$&$3/4$\cr
\raisebox{0pt}[1pt][0pt]{4}&$63/2$&$436/10$&$513/10$&$5/2$\cr
\raisebox{0pt}[1pt][0pt]{5}&$83.189$&$106.44$&$125.54$&$5.792$\cr
\raisebox{0pt}[1pt][0pt]{6}&$176.78$&$217.07$&$261.52$&$459/40$\cr
\raisebox{0pt}[1pt][0pt]{7}&$339.51$&$405.5$&$497.62$&$20.834$\cr
\raisebox{0pt}[1pt][0pt]{8}&$-$&$-$&$894.45$&$35.781$\cr
\raisebox{0pt}[1pt][0pt]{9}&$-$&$-$&$1544.9$&$59.135$\cr
\end{tabular}
\end{center}
\end{table}
}

\vbox{
\begin{table}[hbt]
\caption{The maximal supercurrent  ($\phi\approx\pi$)
in units $I_{\rm res}^{(0)}(\phi)$ for relatively long arrays and
strong coupling. For different bases the results were obtained 
by diagonalisation and the renormalised value is for third
order renormalisation and $2N(N-1)$-state $P$-space.
\label{tab:renor}}
\vspace{0.2cm}
\begin{center}
\footnotesize
\begin{tabular}{cccccc}
{\raisebox{0pt}[5pt][5pt]{$N$}}&$\varepsilon_{\rm J}$&
a-basis&b-basis&c-basis&renorm\cr
\hline
\raisebox{0pt}[7pt][1pt]{6}&$0.1$&$1.485$&$1.488$&$1.491$&$1.492$\cr
\raisebox{0pt}[0pt][1pt]{6}&$0.2$&$1.881$&$1.897$&$1.913$&$1.914$\cr
\raisebox{0pt}[0pt][1pt]{7}&$0.1$&$1.589$&$1.591$&$1.596$&$1.601$\cr
\raisebox{0pt}[0pt][1pt]{7}&$0.2$&$2.072$&$2.087$&$2.113$&$2.117$\cr
\raisebox{0pt}[0pt][1pt]{8}&$0.1$&$1.689$&$1.691$&$1.697$&$1.708$\cr
\raisebox{0pt}[0pt][1pt]{8}&$0.2$&$2.256$&$2.271$&$2.303$&$2.316$\cr
\raisebox{0pt}[0pt][1pt]{9}&$0.1$&$1.786$&$1.788$&$-$&$1.815$\cr
\raisebox{0pt}[0pt][1pt]{9}&$0.2$&$2.435$&$2.448$&$-$&$2.508$\cr
\raisebox{0pt}[0pt][1pt]{10}&$0.1$&$1.881$&$1.883$&$-$&$1.923$\cr
\raisebox{0pt}[0pt][1pt]{10}&$0.2$&$2.601$&$2.621$&$-$&$2.690$\cr
\end{tabular}
\end{center}
\end{table}
}

\vbox{
\begin{table}[hbt]
\caption{The ratios $W_{\rm non}$ corresponding to
nonideal gating in an inhomogeneous array. The 
numerical values $W_{\rm non,num}$ have 
obtained by numerical integration for $N=5$ a-basis with
$\varepsilon_{\rm J}=0.04$ or b-basis with $\varepsilon_{\rm J}
=0.03$. The values of $W_{\rm non,ren}$ and $W_{\rm non,prod}$
are obtained by renormalisation when inhomogeneity and nonideal gating
sequences are treated simultaneously and separately, respectively.
\label{tab:inhonon}}
\vspace{0.2cm}
\begin{center}
\footnotesize
\begin{tabular}{ccccc}
{\raisebox{0pt}[5pt][5pt]{$X_{\rm inh}$}}&$X_{\rm non}$&
$W_{\rm non,num}$&$W_{\rm non,ren}$&$W_{\rm non,prod}$\cr
\hline
\raisebox{0pt}[7pt][1pt]{$0.0131$}&$0.0377$&$1.0178$&$1.018$&$1.018$\cr
\raisebox{0pt}[0pt][1pt]{$0.0263$}&$0.0130$&$1.0082$&$1.0085$&$1.0085$\cr
\raisebox{0pt}[0pt][1pt]{$0.0292$}&$0.0173$&$1.0095$&$1.0115$&$1.0115$\cr
\raisebox{0pt}[0pt][1pt]{$0.0387$}&$0.0163$&$1.0169$&$1.0173$&$1.0173$\cr
\raisebox{0pt}[0pt][1pt]{$0.0515$}&$0.0174$&$1.0277$&$1.0287$&$1.0289$\cr
\raisebox{0pt}[0pt][1pt]{$0.0541$}&$0.0244$&$1.0339$&$1.0346$&$1.0349$\cr
\raisebox{0pt}[0pt][1pt]{$0.0653$}&$0.0237$&$1.0463$&$1.048$&$1.0477$\cr
\raisebox{0pt}[0pt][1pt]{$0.0741$}&$0.0201$&$1.0572$&$1.0577$&$1.058$\cr
\raisebox{0pt}[0pt][1pt]{$0.0783$}&$0.0315$&$1.0697$&$1.0706$&$1.0714$\cr
\end{tabular}
\end{center}
\end{table}
}

\vbox{
\begin{figure}
\begin{center}
      \mbox
      {\epsfig{file=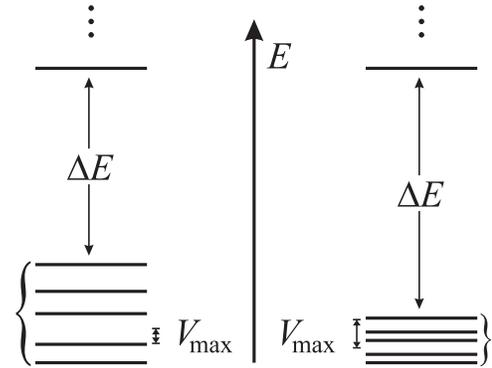,height=48truemm}
      }
\end{center}
\caption{Schematic view of two possible 5-state dominant
systems. On the right-hand-side  all low-lying levels are closely
packed in energy while on the left-hand-side the spread is
large as compared to $V_{\rm max}$. In both cases the
requirements for few-state dominance are well satisfied.
\label{fig:first}
}
\end{figure}
}

\vbox{
\begin{figure}
\begin{center}
      \mbox
      {\epsfig{file=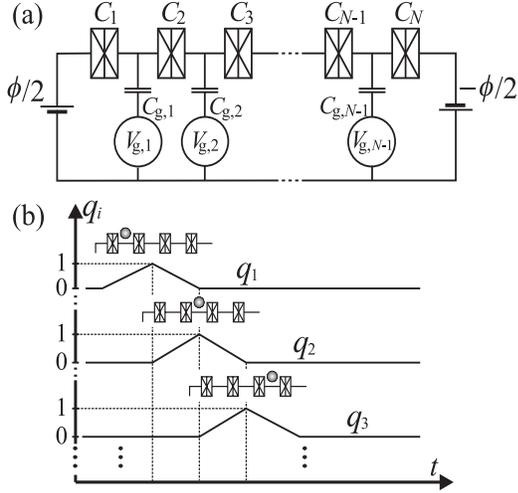,height=65truemm,width=68truemm}
      }
\end{center}
\caption{(a) A schematic drawing of a gated Josephson array of
$N$ junctions. In pumping Cooper pairs, gate voltages $V_{{\rm g},k}$ are
operated cyclically. $C_k$ are the capacitances of the
junctions and $C_{{\rm g},k}$ are the gate capacitances.  b)
A train of gate voltages to carry a charge in a pump. Here $q_k=
-C_{{\rm g},k}V_{{\rm g},k}/2e$. The dominant state at the
turning points of gate voltages are also shown.
\label{fig:cppump}
}
\end{figure}
}

\vbox{
\begin{figure}
\begin{center}
      \mbox
      {\epsfig{file=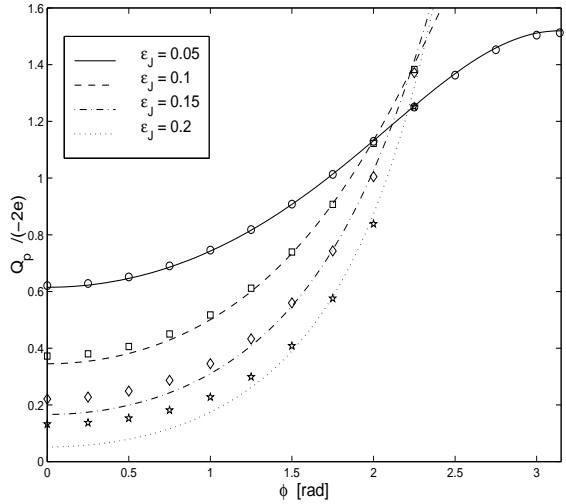,height=72truemm,width=75truemm}
      }
\end{center}
\caption{The pumped charge $Q_{\rm p}/(-2e)$ as a function of $\phi$
for some values of $\varepsilon_{\rm J}$ and $N=3$. Curves denote
renormalised values and symbols numerical values which were obtained
for a $41$-state basis. Pumped charge is symmetric in $\phi$ and its
period is $2\pi$.
\label{fig:n3inacc}
}
\end{figure}
}

\vbox{
\begin{figure}
\begin{center}
      \mbox
      {\epsfig{file=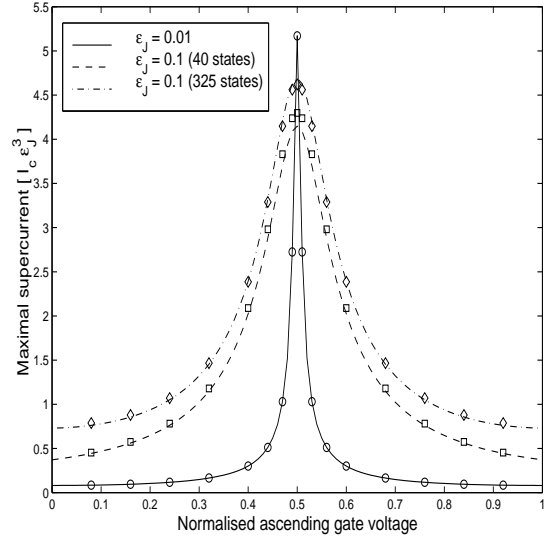,height=72truemm,width=72truemm}
      }
\end{center}
\caption{The maximal value for the supercurrent in units
$I_{\rm C}\varepsilon_{\rm J}^3$ for $N=5$.
Curves denote renormalised values and symbols  numerical values
corresponding to bases with $40$ and $325$ states. 
In case $\varepsilon_{\rm J}=0.01$ the differences
between bases can hardly be seen even at the degeneracy point.
\label{fig:supercur}
}
\end{figure}
}

\vbox{
\begin{figure}
\begin{center}
      \mbox
      {\epsfig{file=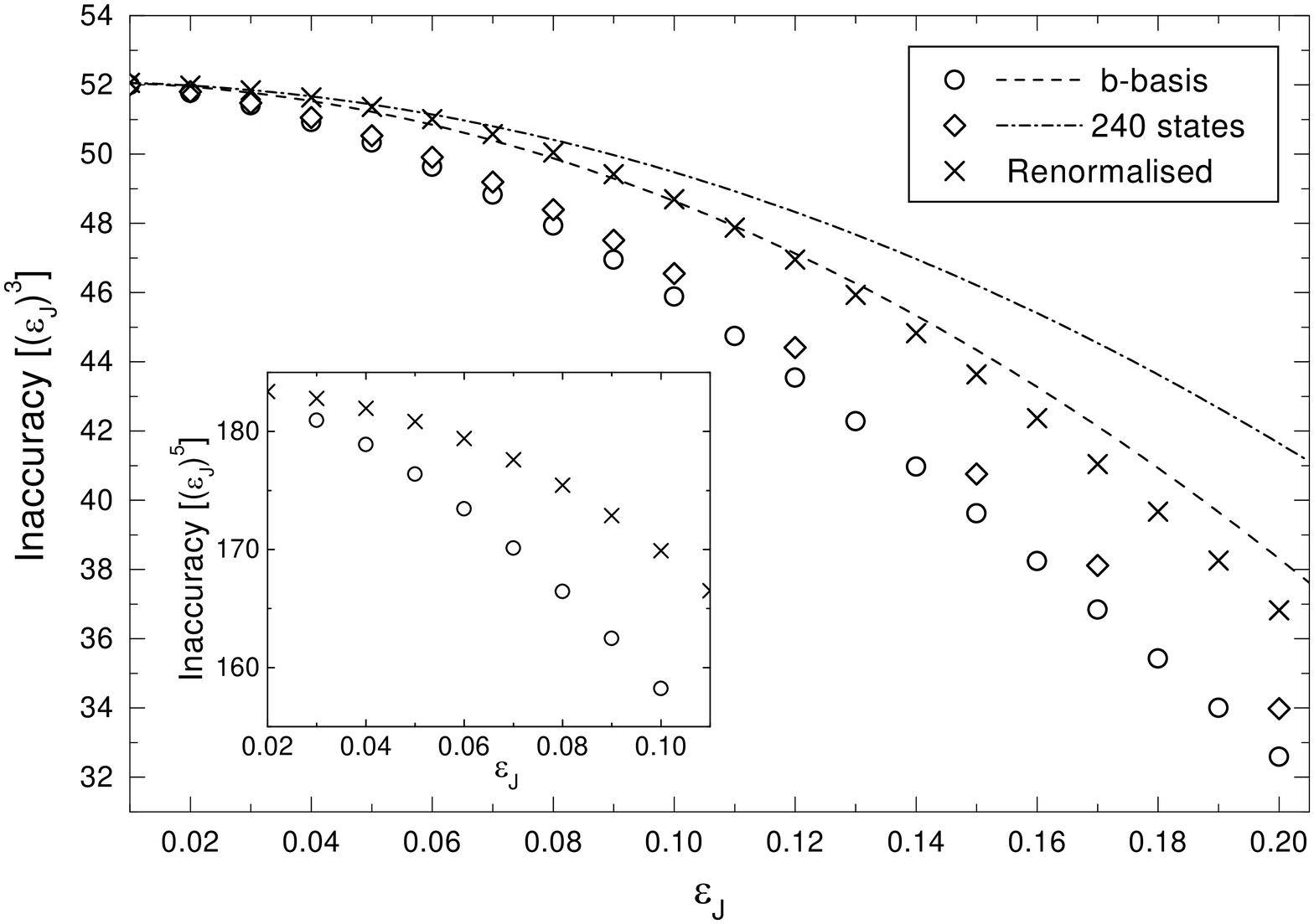,height=58truemm,width=75truemm}
      }
\end{center}
\caption{The inaccuracy of the pumped charge $Q_{\rm p}/(-2e)$ 
as a function of  $\varepsilon_{\rm J}$ for different bases and $N=5$.
Curves denote analytical power expansions and symbols
numerical or renormalised values. The inaccuracy is given in units
$\varepsilon_{\rm J}^3$ and the phase difference used is $\phi=0$.
Inset shows the corresponding results for the $N=7$, 336-state basis. 
\label{fig:n5inacc}}

\end{figure}
}

\vbox{
\begin{figure}
\begin{center}
      \mbox
      {\epsfig{file=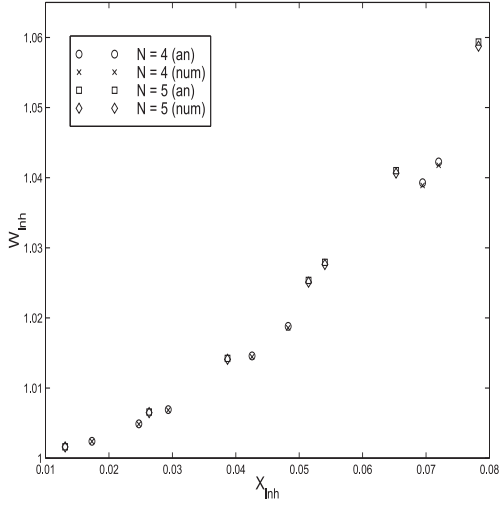,height=72truemm,width=75truemm}
      }
\end{center}
\caption{The ratios $W_{\rm inh}$ 
between the inhomogeneous and the homogeneous
inaccuracies from analytical expression (\ref{inhinaccu})
and numerical calculations as
functions of the inhomogeneity index
$X_{\rm inh}$ which is defined in Eq.~(\ref{inhomogene}).
Numerical results were obtained for b-bases with
$\varepsilon_{\rm J}=0.02$ and 
$\varepsilon_{\rm J}=0.03$ for $N=4$ and $N=5$, respectively.
\label{fig:inhinaccu}
}
\end{figure}
}

\vbox{
\begin{figure}
\begin{center}
      \mbox
      {\epsfig{file=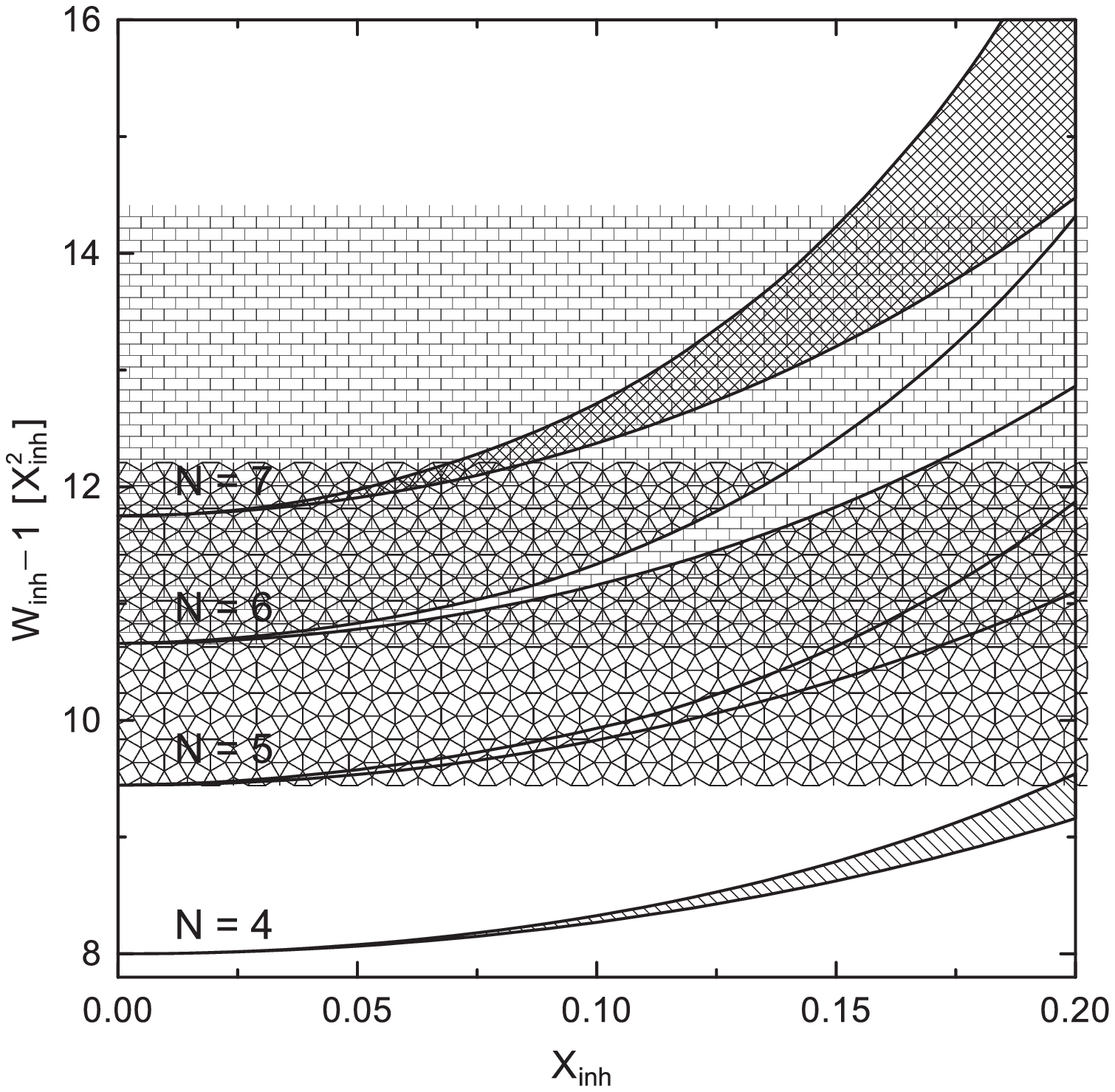,height=70truemm,width=70truemm}
      }
\end{center}
\caption{The limits for the ratio $W_{\rm inh}$ as 
a function of $X_{\rm inh}$ for
array lengths $N=4$ to $N=7$. For small values of $X_{\rm inh}$
$W_{\rm inh}\approx 1+a^{({\rm inh})}_N\cdot X_{\rm inh}^2$ which allows
several cases to be presented simultaneously.
\label{fig:inhlimits}
}
\end{figure}
}

\vbox{
\begin{figure}
\begin{center}
      \mbox
      {\epsfig{file=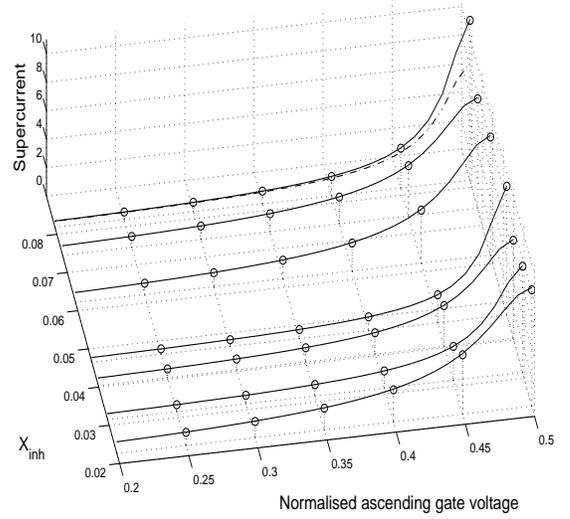,height=70truemm,width=73truemm}
      }
\end{center}
\caption{A three-dimensional plot of the maximal supercurrent for
$N=6$ c-basis in units $I_{\rm c}\varepsilon_{\rm J}^4$ for several
sets $\{c_k\}$ corresponding to different $X_{\rm inh}$.  The gate
voltages are chosen from the first leg of the saw-tooth gating
path. Junction capacitances have been chosen randomly as well as the
ratios $\varepsilon_{\rm J}$ which lie between 0.02 and 0.06.  Solid
curves denote analytical values and discrete symbols numerical values.
The modifications of the supercurrent are well reproduced even for
larger inhomogeneities. The dash-dot curve represents the homogeneous
supercurrent the largest $X_{\rm inh}$..
\label{fig:inhosup}}
\end{figure}
}

\vbox{
\begin{figure}
\begin{center}
      \mbox
      {\epsfig{file=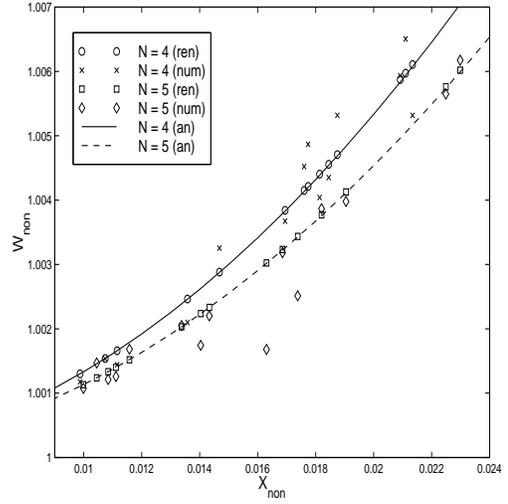,height=72truemm,width=75truemm}
      }
\end{center}
\caption{$W_{\rm non}$ for homogeneous arrays and nonideal
gating sequences as function of $X_{\rm non}$. The renormalised
values are almost identical to the asymptotical
expansions (\ref{eq:nonidlim}) shown as lines.
Numerical values agree reasonably well with theoretical results.
We used a-bases with $\varepsilon_{\rm J}=0.05$ and
$\varepsilon_{\rm J}=0.04$ for $N=4$ and $N=5$, respectively.
\label{fig:nonideal}}
\end{figure}
}


\begin{references}
\bibitem{end:biasing}The charging energy differences in the presence
of an external bias are unaffected for a homogeneous array. The
charging energy differences only deviate from those given by
(\ref{generchar0}) for an inhomogeneous array with an external bias.
\bibitem{note1}The $240$-state basis and some of the data
 was created when the authors worked on Ref. 4. Since calculations are
take a lot of CPU-time and the value of $b_{-1}^{(N)}$ for this basis
is just 2 \% short of the full c-basis value we chose to fill the
missing data rather than adopt the c-basis.
\bibitem{note3}Our calculations show that the inaccuracies
can be predicted even on a ``leg by leg''-basis but since
legwise inaccuracies can not be measured they can not be
considered as results. In addition, the agreement is better for
the full cycle.
\bibitem{note4}The lower limit in Fig.~\ref{fig:inhlimits} 
is not valid in the strict, mathematical sense but the violation
is extremely weak and it only occurs for $X_{\rm inh}\le 0.03$.
The asymptotical limit (\ref{eq:asymplim}) is never violated,
though.
\end{references}
\end{document}